\documentclass[preprint,3p,twocolumn]{elsarticle}
\usepackage{geometry}
\usepackage{fancyhdr}

\biboptions{numbers,sort&compress}
\journal{Computational Materials Science}

\usepackage{amsmath}
\usepackage{amssymb}
\usepackage{multirow}
\usepackage{lineno}
\biboptions{comma,square,sort&compress}

\begin{document}
%%%%%%%%%%%%%%%%%%%%%%%

\begin{frontmatter}

\title{Multiscale simulation of powder-bed fusion processing of metallic alloys}

\author[imdea,indus]{S.M.~Elahi}
\author[imdea]{R.~Tavakoli}
\author[imdea]{A.K.~Boukellal}
\author[imdea,caminos]{T.~Isensee}
\author[imdea,indus]{I.~Romero}
\author[imdea]{D.~Tourret}

\address[imdea]{IMDEA Materials, Madrid, Spain}
\address[indus]{Universidad Polit\'ecnica de Madrid, E.T.S. de Ingenieros Industriales, Madrid, Spain}
\address[caminos]{Universidad Polit\'ecnica de Madrid, E.T.S. de Ingenieros de Caminos, Madrid, Spain}

% \cortext[corr]{Corresponding author; Email address: damien.tourret@imdea.org}

\begin{abstract}

We present a computational framework for the simulations of powder-bed fusion of metallic alloys, which combines: (1) CalPhaD calculations of temperature-dependent alloy properties and phase diagrams, (2) macroscale finite element (FE) thermal simulations of the material addition and fusion, and (3) microscopic phase-field (PF) simulations of solidification in the melt pool.  The methodology is applied to simulate the selective laser melting (SLM) of an Inconel 718 alloy using realistic processing parameters.  We discuss the effect of temperature-dependent properties and the importance of accounting for different properties between the powder bed and the dense material in the macroscale thermal simulations.  Using a two-dimensional longitudinal slice of the thermal field calculated via FE simulations, we perform an appropriately-converged PF solidification simulation at the scale of the entire melt pool, resulting in a calculation with over one billion grid points, yet performed on a single cluster node with eight graphics processing units (GPUs).  These microscale simulations provide new insight into the grain texture selection via polycrystalline growth competition under realistic SLM conditions, with a level of detail down to individual dendrites.

\end{abstract}


%%%%%%%%%%%%%%%%%%%%%%%
\end{frontmatter}
%%%%%%%%%%%%%%%%%%%%%%%

\thispagestyle{fancy}

\section{Introduction}
\label{intro}

Fusion-based additive manufacturing (AM) processes for metals combine a formidable breadth of
interdependent physical phenomena \cite{debroy2018additive}.  In this context, Integrated
Computational Materials Engineering (ICME) \cite{kuehmann2009computational, allison2011integrated,
panchal2013key, xiong2015integrated} offers a prominent pathway to accelerate the design of novel alloys
and to optimize manufacturing processes, by reducing the dependence on experimental trial-and-error
iterations.  ICME heavily relies on linking multiple modeling techniques relevant to distinct
length/time scales and/or different physical phenomena.  Yet, despite the fact that a wide range of models has
been established, the efficient bridging between models and between scales remains a major
challenge.

The rapid deployment of metal additive manufacturing in the last decade has led to the development of many modeling approaches for powder-bed fusion processes \cite{francois2017modeling, korner2020modeling, debroy2021metallurgy}.
At the macroscopic/process scale, multi-physics models have been developed in order to combine fluid and solid mechanics and thermal transport (conduction, convection, and radiation) during melting and solidification phenomena resulting from the moving heat source (e.g. laser or electron beam) above a powder bed \cite{otto2010towards, khairallah2016laser, panwisawas2017keyhole, yan2017multi, zhang2018numerical, tang2020physics}. 
These studies have demonstrated the importance of fluid flow within the melt pool --- in particular, Marangoni convection and recoil pressure --- on the formation of critical defects, such as porosity, spattering, denudation, and balling \cite{khairallah2016laser, panwisawas2017keyhole, yan2017multi, zhang2018numerical, tang2020physics}. 
The effect of the scanning strategy, e.g. the hatch pattern \cite{yan2017multi, zhang2018numerical}, was also shown to have a key effect on the quality of the build. 
Scanning path, cross-section thickness, and the presence of additional lasers were also found to have a strong influence on residual stresses, which also exhibit significant differences between vertical (build) and horizontal directions \cite{zhang2018numerical}. 
Meso-scale phase-field (PF) models of the melting and solidification of powder particles were also shown to capture the effect of the heat source power and speed on the densification, defects, and surface morphology \cite{lu2018phase, yang20193d, yang2021phase}. 
Three-dimensional PF simulations of polycrystalline grain structures at the melt pool scale were recently reported in the case of high growth velocities reaching absolute interfacial stability \cite{chadwick2021development}. 
However, it remains limited to alloys and growth velocities that yield complete solute trapping and planar re-stabilization of the solid-liquid interface. As most metallic alloys develop dendritic or cellular microstructures during AM, grain growth competition typically involves complex mechanisms such as dendritic sidebranching and impingement of individual dendrite or cells \cite{tourret2015growth, tourret2017grain}. 

At the scale of the grain structure, the choice of the model often results
in the classical trade-off between efficiency (and hence scale) and physics-based accuracy.
Among coarse-grained, efficient approaches, those based on the kinetic Monte Carlo (KMC) method
conveniently allow three-dimensional simulation at the full melt pool scale, including multiple
layers \cite{rodgers2017simulation, wei2017crystal}. 
However, they usually do not integrate some essential physics behind polycrystalline microstructure selection
(e.g. the anisotropic growth kinetics and preferred growth directions of crystalline grains).
Models based on meso-scale cellular automata (CA) coupled with macro-scale thermomechanics (e.g. using finite elements or differences) provide a good compromise between efficiency and
physics-based considerations \cite{rappaz1993probabilistic, gandin1994coupled, chen2016three, rai2016coupled,
koepf2019numerical, lian2019cellular, mohebbi2020implementation}.  In such models, grains can be
constructed from polyhedral building blocks, whose vertices mark the preferred crystallographic
growth directions.  Growth velocities in these directions follow simplified, yet physics-based,
kinetic laws for crystal growth --- e.g. using Ivantsov-based relations or power laws relying on local
supersaturation or undercooling.  As such, these models still include some adjustable
phenomenological parameters.  It was recently shown that polycrystalline growth can be predicted by
CA-based models with an accuracy comparable to that of phase-field, as long as the cell size is
adjusted to the length scale relevant to the grain growth competition --- e.g. the ``height''
difference between the two competing grains \cite{pineau2018growth} or the spacing between active
secondary branches \cite{dorari2021growth}.  

At the scale of dendritic/cellular arrays, most-accurate physics-based models require a numerical discretization commensurate
with the microstructural scale of interest, and therefore result in computationally expensive
simulations.  Phase-field (PF) models are arguably the most accurate approach to simulate the
evolution of morphologically complex interfaces based solely on thermodynamics and kinetics
considerations \cite{boettinger2002phase, chen2002phase, moelans2008introduction,
steinbach2009phase, steinbach2013phase, tourret2022phase}.   The method has been particularly
successful in the field of solidification, often using ``mesoscale'' interface formulations that
remain faithful to the well-known sharp-interface problem even for diffuse interfaces much wider
than the actual interface width or its capillary length \cite{boettinger2002phase,
steinbach2013phase}.  However, spatial discretization remains upward-bounded by the typical
microstructural length scale, i.e. the local interface curvature \cite{karma1998quantitative, echebarria2004quantitative, steinbach2009phase}.  Since
dendrite tip radii can go down to a few tens of nanometers in fusion-based metal AM, this spatial
discretization requirement makes quantitative simulations computationally demanding.
Therefore, reported studies that have used PF to simulate solidification within an AM melt pool have
been limited in size \cite{keller2017application,
boussinot2019strongly, kundin2019microstructure, pinomaa2020significance, karayagiz2020finite}.

Computational approaches have been proposed that combine macroscopic thermal models with lower scale
microstructure models \cite{francois2017modeling, korner2020modeling, debroy2021metallurgy}.  
Recent studies have used macroscopic thermal simulation of AM processing, in order to provide thermal conditions for lower scale PF simulations of solidification \cite{fallah2012phase, keller2017application, karayagiz2020finite, berry2021toward}.
Yet, while full melt pool simulations appear within reach using computationally-efficient parallelized implementations (see, e.g. \cite{yu2019evolution}), PF simulations have been mostly restricted to the growth of a handful of dendrites with thermal conditions relevant to a subset of the melt pool region \cite{fallah2012phase, keller2017application, karayagiz2020finite, berry2021toward}.

In this article, we present a multiscale modeling framework for powder-bed fusion processes, and
demonstrate its capabilities focusing on the simulation of selective laser melting of Nickel-based
superalloy Inconel 718.  
The modeling approach does not focuses on melt pool fluid dynamics nor defect formation.
Rather, a key objective is to obtain a sensible physics-based prediction of the key
microstructural features (e.g. grain sizes, dendritic spacings, and chemical segregation between and
within the grains) at the scale of the entire melt pool.  The computational framework combines three
main components.  First, the CalPhaD method provides thermophysical properties of the alloy and its
phase diagram.  Then, a three-dimensional macroscale finite element (FE) model is used to assess the
temperature field, with particular emphasis on the accurate description of the melt pool size and
shape.  Finally, using the FE-calculated temperature field, a quantitative phase-field (PF)
simulation is performed along a two-dimensional slice (here using the most computationally demanding
longitudinal slice) at the scale of the entire melt pool.  By linking the models with one another,
we ensure that microstructure simulations are performed using realistic alloy and process parameters
--- and importantly the actual resulting melt pool size and shape.

%=================================================================

\section{Methods}
\label{method}

\begin{figure*}[t!]
    \centering
    \includegraphics[width=1.6\columnwidth]{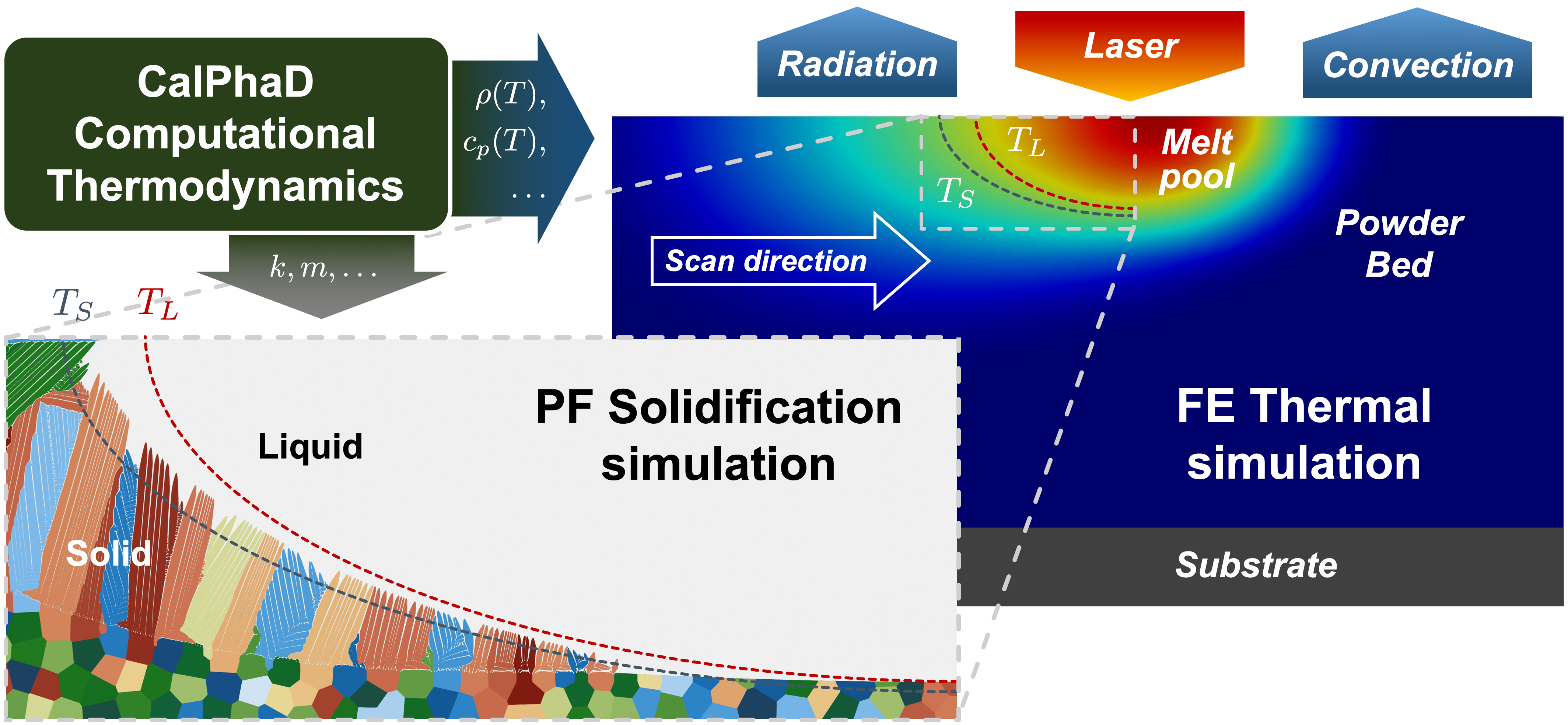}
    \caption{ Main components of the multiscale modeling strategy, namely: CalPhaD thermodynamics, Finite Elements thermal simulation of SLM processing, and Phase-Field solidification simulation at melt pool scale.
    }
    \label{fig:schematic}
\end{figure*}

The multiscale simulation approach presented here relies on three major components, namely: (1)
computational thermodynamics (CalPhaD) to calculate temperature-dependent properties of complex
alloys and their phase diagram, (2) macroscale (FE) thermal simulation of the powder-bed fusion, and (3) microscale (PF)
simulation of microstructure development by crystal growth within the melt pool.  Macro- and
microscale simulations are coupled through the temperature field, estimated within a two-dimensional
slice in the vicinity of the melt pool, and used as input to the phase-field simulations.  The
latter are upscaled using massive parallelization on Graphics Processing Units (GPUs) in order to
make them applicable at the full melt pool scale, without compromising their accuracy,
i.e. retaining the required grid size of a few nanometers.  The key features of the resulting
computational framework are schematized in Figure~\ref{fig:schematic} and details are presented in the
following subsections.

\subsection{Computational thermodynamics of multicomponent alloy}
\label{method:calphad}

Thermodynamic properties used in the FE and PF simulations are calculated using the CalPhaD
(CALculation of PHAse Diagrams) method.  The method is based on the global minimization of the free
energy using databases describing the free energies of individual phases.  Extensive databases thus
allow calculating thermodynamic equilibria, i.e phase diagrams, for complex multicomponent
alloys, and therefore to study the effect of individual alloying element variation.

\subsubsection{Temperature-dependent alloy properties}
\label{method:calphad:fem}

We calculated alloy properties as a function of temperature assuming complete equilibrium, i.e. a
lever rule solidification path, but the methodology would be just as readily applicable considering
alternative assumptions, e.g. Gulliver-Scheil \cite{kurz1984fundamentals}.  
In particular, within a temperature range spanning
all phase transformations, here from 250\,K to 5000\,K, we estimated the temperature-dependent
density, $\rho (T)$, as well as the enthalpy per unit mass, $h(T)$.  From the latter, we calculated
an effective heat capacity $c_p(T)=\partial h(T)/\partial T$ that includes not only the actual heat
capacity but also the enthalpies (latent heat) of transformations in the relevant temperature
ranges.  From the tabulation of the fraction of liquid versus temperature, one can also extract the
key transformation temperatures, namely liquidus ($T_L$), solidus ($T_S$) and boiling ($T_V$)
temperatures.

In particular, we apply the approach to a multicomponent Inconel 718 (IN718) superalloy, of
composition listed in Table~\ref{tab:alloy}, using the software ThermoCalc with Ni-alloys database
TCNI8.   In the interest of computational efficiency, the resulting $h(T)$ and $\rho(T)$ used in the
FE calculations are approximated by piecewise linear functions (see Section~\ref{resu:calphad}).
CalPhaD-calculated properties can also be complemented by any further temperature-dependent property
from other sources, such as illustrated here using a thermal conductivity, $\kappa(T)$, from the
literature \cite{mills2002recommended}.

\begin{table}[t]
\centering
\caption{Chemical composition of superalloy Inconel 718}
\begin{tabular}{l l l l l l l}
\hline
Element &  Ni &  Cr  &  Nb &  Mo &  Ti &  Al \\
\hline
Weight \% &  50.0 &  17.0  &  4.75 &  2.8 &  0.65 &  0.2 \\
\hline
\end{tabular}
\label{tab:alloy}
\end{table}

\subsubsection{Phase diagram and pseudo-binary approximation}
\label{method:calphad:pf}

CalPhaD is also used to calculate phase diagram features for the PF simulations.  As described in
section~\ref{method:pf}, we use a reference quantitative phase-field model for binary alloy
solidification.  Therefore, the alloy is approximated by a pseudo-binary Ni-5wt\%Nb, but the coupling approach
could be readily extended without loss of generality to further models, e.g. using models for
multicomponent \cite{eiken2006multiphase, nestler2011phase, ohno2012quantitative} and/or
incorporating rapid solidification effects \cite{steinbach2012phase,  zhang2012phase,
kundin2015phase, pinomaa2019quantitative, kavousi2021quantitative, kim2021phase}.  Solute trapping
is intentionally left out of the current scope, such that the solid-liquid interface is assumed to
remain close to thermodynamic equilibrium, with local deviation from the phase diagram solely due to
solute and curvature effects.  Consequently, to keep simulations quantitative, applications of the
methodology are limited to a scan velocity at which the solute trapping is negligible
(section~\ref{method:simu}).  We focus specifically on Niobium because it is one of the solute
elements in IN718 that exhibits the highest segregation during solidification, thus playing a key
role in the formation of secondary intermetallic phases as well as in the hot cracking susceptibility of highly segregated grains boundaries.

There are several different ways to approximate of a multicomponent alloy as a pseudo-binary system.
The most straightforward route is to directly consider the actual binary Ni-Nb system at the
relevant Nb concentration (see, e.g., \cite{keller2017application}).  However, in the scope of a
coupling of thermal fields while using thermophysical properties of the full alloy, this may lead to
an important mismatch in transformation temperatures (e.g. $T_L$, $T_S$) between FE (full alloy) and PF (pseudo-binary) calculations.  
Instead, here we aim at closely matching the liquidus temperature $T_L$, as well as
the solute partitioning of Nb at the solid-liquid interface for a temperature close to $T_L$.  
To do
so, we use CalPhaD (ThermoCalc, TCNI8) to compute the thermodynamic equilibrium of the complete alloy (Table~\ref{tab:alloy}) at $T=T_L$.
At this point, we estimate the solute partition coefficient of Niobium, $k=c_s/c_l$, with $c_s$ and $c_l$ the respective concentration (weight) of Nb in the solid and liquid phase, and the liquidus slope with respect to the Nb concentration, $m=\partial T_L/\partial c$.
Then, we calculate the corresponding (fictitious) pure solvent melting temperature, $T_M'$, and the solidus temperature, $T_S'$, assuming a linearized phase diagram, i.e. using $T_L=T_M'+mc_\infty$ and
$T_S'=T_M'+(m/k)c_\infty$, with $c_\infty=5$wt\%Nb the nominal alloy solute concentration considered in the pseudo-binary approximation.  
(Here, prime symbols on $T_M'$ and $T_S'$ denote this pseudo-binary approximation.)
Hence, this method
may lead to some discrepancy in terms of the linearized $T_M'$ and $T_S'$ (see
section~\ref{resu:calphad}) but it has the key advantage of retaining an accurate description of
solute partitioning and interface equilibrium in the vicinity of $T=T_L$ as well as a matching of
temperature $T_L$.  Other parameters, such as the solid-liquid interface Gibbs-Thomson coefficient
or the liquid solute diffusivity are extracted or calculated from literature data (see
section~\ref{method:simu:pf} and Table~\ref{tab:param:pf}).

\subsection{Macroscopic thermal simulations}
\label{method:fem}

We consider the thermal problem of a moving heat source above a powder bed. 
Neglecting mechanics, in particular fluid flow, is expected to have a strong influence.
However, for printing in conduction mode, a thermal model is expected to provide a reasonable estimate of the temperature profile in and around the melt pool.

\subsubsection{Thermal problem}
\label{method:fem:eqs}

Considering a bounded domain $\Omega$ in $\mathbb{R}^3$, the governing equation for the heat transfer problem can be written as
\begin{equation}
       \rho(T)\, c_{p}(T)\, \dot{T}(\mathbf x) = \nabla \cdot \big( \kappa(T)\nabla T \big), \quad  \mathbf x \in \Omega
    \label{eq_thermal}
\end{equation}
where $T$ denotes the temperature, $\rho$ the density, $c_{p}$ the specific heat at constant pressure, $\kappa$ the scalar (i.e. isotropic) conductivity of the material. 
Material properties, namely $\rho$, $c_p$, and $\kappa$, are temperature-dependent, and hence a function of the location $\mathbf x$ and time $t$.

Equation~\eqref{eq_thermal} does not have any volumic heat source term, but the laser beam is modeled by a 2D heat flux applied on the top surface of the powder bed.  A standard Gaussian model is used that assumes a heat input
that is symmetric with respect to the laser beam axis and total irradiance
\begin{equation}
    I = \frac{2AP}{\pi w^2}\exp\left(-\frac{2r^2}{w^2}\right)
    \label{eq1}
  \end{equation}
where $A$ is the absorptivity of the powder bed, $P$ is the laser power, $w$ is the laser beam diameter, and $r$ is the radial distance from the center of the laser beam.

Along external boundaries, convection and radiation are considered. The convective heat transfer is modeled as
\begin{equation}
    q_{c}(T) = h_c (T - T_{\rm ext})
    \label{eq4}
\end{equation}
where $h_c$ is the convective heat transfer coefficient and $T_{\rm ext}$ is the temperature of the surrounding environment. Due to the high temperatures reached during laser melting, radiation is an important heat loss mechanism, which is modeled using the Stefan-Boltzmann law, with the radiative heat flux expressed as
\begin{equation}
    q_{r}(T) = \sigma \epsilon (T^4 - T_{\rm ext} ^4)
    \label{eq5}
\end{equation}
where $\sigma$ is the Stefan-Boltzmann constant and $\epsilon$ is the material emissivity.
Finally, at the bottom of the domain, a substrate material is considered to exchange heat via conduction with the powder bed and the solidified material, and a Dirichlet boundary condition of the form $T=T_{\rm ext}$ is set at the bottom of the substrate.

%=========================================

\subsubsection{Material addition, phase change, and resulting properties}
\label{method:fem:slm}

The material above the substrate can take three different ``states'': Powder, Solid, and Fluid.
The ``powder'' state is actually used for the powder bed, which is a combination of powder particles and gas, with homogenized properties reflecting that of the solid, gas, and morphological descriptors of the powder bed (e.g. average particle size, packing factor, etc.) as described in the following paragraphs.
After the activation of a new layer (see section~\ref{method:fem:implem}), the initial state of the
added material is powder.  At each time step, when the temperature of an element exceeds the alloy
liquidus temperature, $T_L$, a transition occurs from powder to fluid state.  
Subsequently, the state of a fluid material point whose temperature becomes lower than the solidus temperature, $T_S$, is switched from fluid to solid.

The state of a point can change several times between fluid and solid depending on the local thermal
history.  However, once an element has switched from its initial powder state to a dense state, its
properties remain those of a dense material, for the rest of the simulation.  The subsequent changes
between solid and fluid states are intrinsically represented by the temperature-dependent properties
via their variations upon phase transformation (see Section~\ref{resu:calphad} and
Figure~\ref{fig:calphad}).  The latent heat of transformations are not explicitly introduced in
Eq.~\eqref{eq_thermal}.  Instead, the model uses an effective heat capacity, calculated as
$c_p(T)=\partial h(T)/\partial T$ for the entire temperature range, which therefore incorporates the
effect of these transformations. 
In practice, we use piecewise linear fits of the CalPhaD calculated $h(T)$ and $\rho(T)$ and we use the 
high-slope region of $h(T)$ just above $T_V$ for any temperature $T\geq T_V$ (see later Figure~\ref{fig:calphad}).
This way, the critical effect of evaporative cooling \cite{karayagiz2019numerical} is incorporated phenomenologically 
and the temperature saturates naturally when $T$ exceeds $T_V$, without artificial increase of thermal conductivity \cite{loh2015numerical} 
or reduction of the heat source term \cite{ivekovic2021crack} in the vicinity of the evaporation temperature.

Temperature-dependent properties of the dense material are those calculated from CalPhaD or
extracted from the literature.  Properties of the powder bed are evaluated from those of the dense
material, the surrounding gas, as well as additional descriptors of the powder morphology and
packing. The density of the powder bed, $\rho_p$, is estimated using a classical rule of mixture
\begin{equation}
       \rho_p = (1-\xi)\rho_s + \xi \rho_g
    \label{eq:powderrho}
\end{equation}
where $\xi$ is the porosity of the powder bed, and $\rho_s$ and $\rho_g$ are densities of bulk material and gas atmosphere (here: Argon), respectively.
For the thermal conductivity of the powder bed, we use the model proposed by Sih and Barlow \cite{sih2004prediction}, derived from the Zehner-Schl\"under-Damk\"ohler equation \cite{damkohler1936einflusse, zehner1970warmeleitfahigkeit} assuming spherical particles and expressed as
\begin{align}
       \frac{\kappa_p}{\kappa_g} = &\left(1-\sqrt{1-\xi}\right) \left(1+\xi \frac{\kappa_r}{\kappa_g}\right) \nonumber\\
        +& \sqrt{1-\xi} \left\{\frac{2}{1-\frac{\kappa_g}{\kappa_s}} \left[\frac{1}{1-\frac{\kappa_g}{\kappa_s}} \ln\left(\frac{\kappa_s}{\kappa_g}\right) - 1\right] + \frac{\kappa_r}{\kappa_g}\right\}
    \label{eq:powderkappa}
\end{align}
where $\kappa_p$, $\kappa_s$, and $\kappa_g$ are conductivity of the powder bed, bulk material, and gas, respectively.
The thermal conductivity due to radiation among particles, $\kappa_r$, is
\begin{equation}
       \kappa_r = 4F\sigma T^3 d
    \label{eq8}
\end{equation}
where $F$ is a view factor, here assumed equal to 1/3 \cite{damkohler1936einflusse, sih2004prediction}, and $d$ is the average diameter of powder particles.

%=========================================

\subsubsection{Numerical implementation}
\label{method:fem:implem}

The thermal model described above was solved numerically in three dimensions using an in-house finite element code (IRIS) and a material library (MUESLI \cite{portillo2017ur}) implemented in C++ programming language.
A Galerkin method employing hexahedral elements was used to discretize the initial boundary value problem in space. The resulting semidiscrete equations were integrated in time with an implicit Backward-Euler method \cite{thomee1997ke}.

Elements for all layers are initially created, including those to be activated at later stages.
They are activated progressively whenever new layers are deposited.
The successive activation of layers is repeated with a latency time between them that either represents the time for new powder application or provides sufficient time for the material to cool down to room temperature.

%=========================================

\subsection{Macroscale to microscale coupling}
\label{method:coupling}

Since thermal diffusivity in metals is several orders of magnitude higher than solute diffusivities, the kinetics of microstructural development within the melt pool is typically limited by the diffusion of species.
Hence, we use the common assumption that the phase transformation does not have a significant effect on the temperature field, such that we can decouple the computation of the temperature field and that of the solidification within the melt pool.
This results in a one-way coupling via the temperature field, which is calculated by FE and then imposed in the PF simulation.

\begin{figure*} [!t]
    \centering
    \includegraphics[width=1.25\columnwidth]{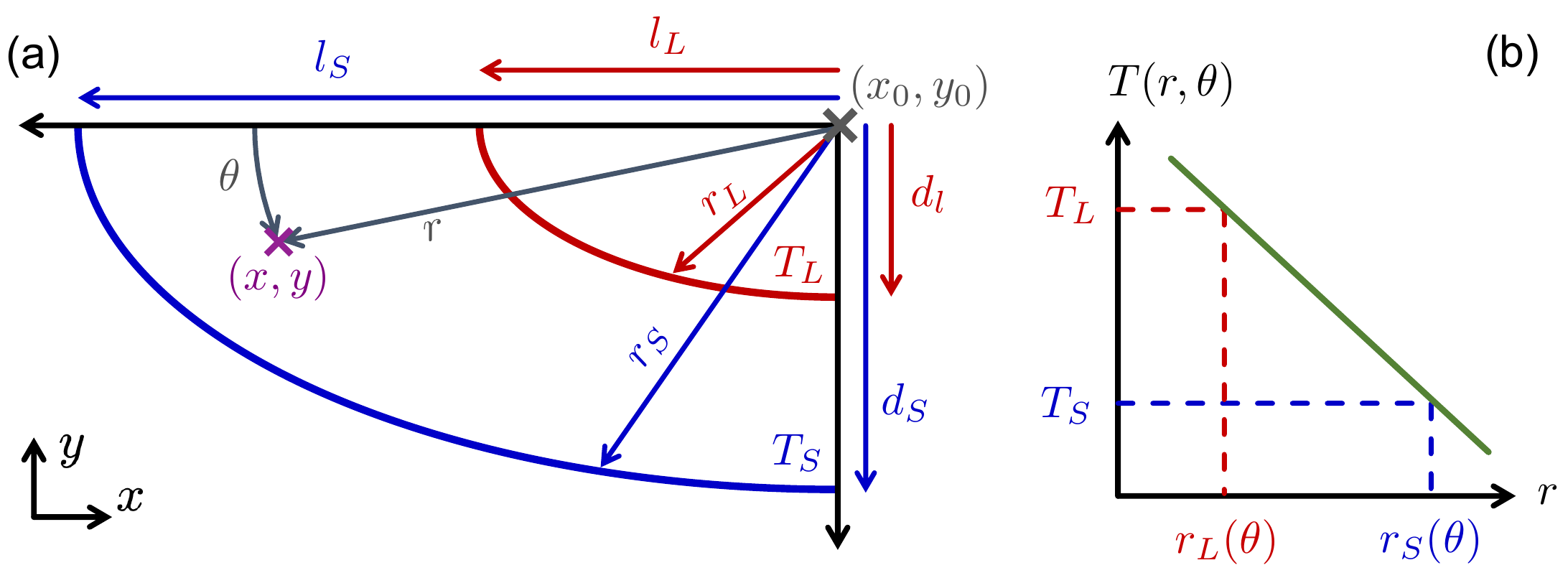}
    \caption{Schematics of the elliptic temperature field approximation.
    }
    \label{fig:elliptic}
\end{figure*}

The resulting PF simulations (see section~\ref{method:pf}) are rigorously analogous to those using the classical one-dimensional frozen temperature approximation \cite{echebarria2004quantitative}, but imposing a different temperature field $T(\mathbf x,t)$.
We extract this 2D temperature field within the longitudinal section of the sample along the laser path.
For the sake of simplicity and computational efficiency, we use an analytical approximation of the temperature field.
The selected expression aims at an approximate yet reasonable description of the temperature field in the region where it matters most for the development of the microstructure, namely between liquidus and solidus temperatures.
Hence, we approximate solidus and liquidus isotherms as two ellipses
\begin{align}
r_L(\theta) & = \sqrt{\frac{(l_L d_L)^2} { \big(d_L \cos
(\theta)\big)^2 + \big( l_L\sin (\theta)\big)^2}}
\label{eq:rl}\\ %
r_S(\theta) & = \sqrt{\frac{(l_S d_S)^2} { \big(d_S \cos
(\theta)\big)^2 + \big( l_S\sin (\theta)\big)^2}}
\label{eq:rs}
\end{align}
where $r_L$ and $r_S$ are the respective radii of the $T=T_L$ and $T=T_S$ ellipses as a function of the angle from the top surface
\begin{align}
\theta &= \tan^{-1} \bigg|\frac{y-y_0}{x-x_0}\bigg|
\end{align}
with $(x_0,y_0)$ the center of the ellipses (see Figure~\ref{fig:elliptic}a).
Here, $x_0$ is the location at which the melt pool is deepest, which may be slightly shifted backwards from the center of the heat source.
Melt pool dimensions appear explicitly in Eqs~\eqref{eq:rl} and \eqref{eq:rs} in the form the length ($l_L$, $l_S$) and depth ($d_L$, $d_S$) of the corresponding solidus (subscript $S$) and liquidus ($L$) isotherms.
The resulting temperature field is then interpolated linearly between $T_L$ and $T_S$, i.e. between $r_L(\theta)$ and $r_S(\theta)$ at a given $\theta$, as (Figure~\ref{fig:elliptic}b)
\begin{align}
T(r, \theta) & = T_L\, + \, (T_0 - T_L)\,
\frac{r-r_L(\theta)}{r_S(\theta)-r_L(\theta)} .
\label{eq:tellip}
\end{align}

One advantage of this expression is that it involves only four adjustable parameters ($l_L$, $l_S$, $d_L$, $d_S$), which can be measured directly and hence monitored automatically from the FE results.
An underlying assumption is that the deepest point of the $T_S$ and $T_L$ isotherms are aligned on the same $x_0$.
However, this is often the case since the two isotherms tend to be close to each other at the bottom of the melt pool (see Figs~\ref{fig:resu:fem3d}-\ref{fig:resu:meltpool}).

Even though we only illustrate the method for the longitudinal cross section with a steady temperature profile moving at constant velocity, the method is readily usable for any cross section or temperature field, including time-dependent temperature fields.
Moreover, one can also easily substitute the proposed analytical expression by an efficient interpolating scheme directly estimating $T(\mathbf x,t)$ from the FE results.

%=========================================

\subsection{Microscale modeling of microstructure growth in the melt pool}
\label{method:pf}

\subsubsection{Phase-field model}
\label{method:pf:eqs}

We consider a classical quantitative PF model for dilute binary alloy solidification \cite{echebarria2004quantitative}.
To reduce the sensitivity of results to spatial grid size, we make use non-linear preconditioning of PF equation \cite{glasner2001nonlinear}.
The final form of PF equations in two spatial dimensions is \cite{tourret2015growth}:

\begin{align}
\nonumber
\bigg(1- \frac{T-T_0}{mc_l^0}\bigg)\, %
a_s(\alpha)^2\, \frac{\partial \psi}{\partial t}   =%
\hskip0.2\textwidth
\\ \nonumber %
\nabla \big[ a_s(\alpha)^2 \big]\cdot \nabla \psi + %
a_s(\alpha)^2\, \bigg[ \nabla^2 \psi\, - \, \phi\, \sqrt{2}\,
|\nabla \psi|^2 \bigg] %
\\ \nonumber  -
\frac{\partial}{\partial x}\, \bigg[ a_s(\alpha)\,
a_s^\prime(\alpha)\, \frac{\partial \psi}{\partial y } \bigg] + %
\frac{\partial}{\partial y}\, \bigg[ a_s(\alpha)\,
a_s^\prime(\alpha)\, \frac{\partial \psi}{\partial x } \bigg] %
\\  +
\sqrt{2}\, \phi - \sqrt{2}\, \lambda (1-\phi^2)\, \bigg(U +
\frac{T-T_0}{mc_l^0(1-k)} \bigg)
\nonumber\\%
\label{eq:P}%
\end{align}
\begin{align}
\nonumber
\bigg(\frac{1+k}{2} - \frac{1-k}{2} \phi\bigg)\, %
\frac{\partial U}{\partial t} =%
\hskip0.2\textwidth
\\ \nonumber %
\nabla \cdot \bigg(\tilde{D}\, \frac{1-\phi}{2}\, \nabla U +
[1+(1-k)\, U]\, \frac{(1-\phi^2)}{4}\, \frac{\partial
\psi}{\partial
t}\, \frac{\nabla \psi}{|\nabla \psi|} \bigg) %
\\  +
[1+(1-k)\, U]\, \frac{(1-\phi^2)}{2\sqrt{2}}\, \frac{\partial
\psi}{\partial t} \nonumber\\%
\label{eq:U}%
\end{align}
where $T$ is the temperature field, $\phi$ is the classical phase-field variable ($+1$ in the solid
and $-1$ in the liquid), $\psi$ is the preconditioned phase-field variable with $\phi(x,y,t)
=\tanh\big(\psi(x,y,t)/\sqrt{2}\big)$, $\alpha = \arctan{\big(\partial_y \psi/\partial_x \psi\big)}$
is the angle between the  solid-liquid interface normal and a fixed reference direction, $U =
\frac{1}{1-k}\, \big(\frac{2\, c/c_l^0}{1\,-\,\phi\, +\, k\,(1\,+\, \phi\, )}\, -\, 1\big)$ is the
dimensionless solute supersaturation, with $c$ the solute concentration field, $c^0_l = c_\infty/k$
the solute concentration of a flat interface at the reference (solidus) temperature $T_0$ for an
alloy of nominal solute concentration $c_\infty$, $k$ is the interface solute partition coefficient
and $m$ is the slope of the liquidus line.  In Eqs~\eqref{eq:P}-\eqref{eq:U}, space is scaled in
units of the diffuse interface width, $W$, and time is in units of the relaxation time, $\tau_0$, at
the temperature $T_0$ \cite{echebarria2004quantitative}. Considering that interpolation functions
used in \eqref{eq:P} and \eqref{eq:U} are determined based on the thin-interface
asymptotic analysis \cite{karma1998quantitative, echebarria2004quantitative}, their solutions will remain quantitative while
using $W$ much larger than the capillarity length. The capillarity length, $d_0$, is expressed at
$T_0$ as $d_0 = \Gamma/\big(|m|c_\infty (1/k - 1)\big)$, where $\Gamma$ denotes the Gibbs-Thomson
coefficient of the solid-liquid interface. The non-dimensional value for the liquid diffusion
coefficient, $\tilde{D}$, and the coupling factor, $\lambda$, are computed according to
\begin{equation}
\label{eq:Dtilde}%
\tilde{D} = \frac{D\tau_0}{W^2} =a_1 \, a_2 \, \frac{W}{d_0}
\end{equation}
\begin{equation}
\label{eq:lambda}%
\lambda = a_1\, \frac{W}{d_0}
\end{equation}
where $D$ is the liquid diffusion coefficient (Eq.~\eqref{eq:U} neglects diffusion in the solid phase), $a_1 =
5\sqrt{2}/8$, and $a_2 = 47/75$. The standard form of the fourfold
anisotropy of the surface tension $\gamma(\bar{\alpha}) =
\bar{\gamma} a_s(\bar{\alpha})$ is considered with
\begin{equation}
\label{eq:as}%
 a_s(\bar{\alpha}) = 1+ \varepsilon_4\, \cos\,(4\bar{\alpha})
\end{equation}
where $\bar{\gamma}$ is the average surface tension in a $(
1 0 0)$ plane, $\varepsilon_4$ is the strength of the surface
tension anisotropy, and $\bar{\alpha}$ is the angle between the
normal to the interface and a fixed crystalline axis. For a
crystal misorientation $\alpha_0$ with respect to the coordinate
axes, the anisotropy as function of the angle $\alpha$ between the
interface normal and the axis $x$ is
\begin{equation}
\label{eq:ast}%
 a_s(\alpha) = 1 + \varepsilon_4\, \cos\,\big(4\, (\alpha - \alpha_0)\big)
\end{equation}
Kinetic undercooling is also ignored, such that $\tau_0$ can be computed as
\begin{equation}
\label{eq:tau0}%
\tau_0 = a_2\, \lambda\, \frac{W^2}{D}
\end{equation}
Thus, $W$ is the only model parameter that should be appropriately
chosen for the purpose of quantitative prediction.

Importantly, the current PF model assumes that the solid-liquid interface is at local equilibrium.
As a result, it is rigorously valid in a regime for which solute trapping effect can be neglected,
and hence is limited to a moderate velocity range toward the lower velocity range relevant to SLM (see Section~\ref{method:simu:proc}).

%=========================================

\subsubsection{Polycrystalline solidification}
\label{method:pf:polyx}

We aim at simulating the epitaxial growth and grain growth
competition of columnar grains with different crystal orientations
in the melt pool, relevant to the process conditions studied here (see Section~\ref{resu:fem}).
A simple method for modeling bi-crystal grain growth
competition \cite{tourret2015growth} is directly extended to polycrystals.
An integer field is used to store the index of grains, which has a value of $-1$ in the liquid
and a positive or zero integer value in each grain.
Each index maps to a given orientation.
Here, for the sake of simplicity, we consider 90 grain orientations,
such that the solid grain index is taken within the range $[0,89]$,
which can be used directly as the value of the grain orientation in degrees.
In the liquid, when $1-\phi^2$ exceeds a certain threshold,
here fixed to $0.01$, the grain index is updated to the index value
most frequently present in the immediate grid point neighborhood.
This method creates a thin halo of orientation field in the liquid around a grain,
thus ensuring the appropriate equation is solved in the vicinity of the interface.
However, when a grain index is attributed to a grid point, the crystal
index field no longer evolves, and the solid-solid grain boundary
will remain ``frozen''. As such, this method does not take into account solid
state grain boundary evolution. Yet, it remains adequate to study grain growth competition
for well developed dendrites, since triple points and grain boundaries are relatively deep,
and the region of interest remains close to the primary tips.
The approach has the key advantage of reducing directly to a reference, thoroughly validated, quantitative PF model  at the solid-liquid interface, while being computationally efficient compared to a model using multiple order parameters (e.g. \cite{steinbach1996phase}).

%=========================================

\subsubsection{Numerical implementation}
\label{method:pf:implem}

Equations \eqref{eq:P} and \eqref{eq:U} are solved in 2D
on a finite-difference grid of square elements of grid
spacing $\Delta x$ using an Euler explicit time scheme with a
constant time step $\Delta t$. The time step size is
taken as $0.3$ times the maximum time step based on the
stability of Laplacian operators in \eqref{eq:P} and \eqref{eq:U}.
The standard second order five-point stencil is used for
discretization of Laplacian operators. Other terms in \eqref{eq:P}
and \eqref{eq:U} are discretized by central difference schemes
(see appendix A and B of \cite{tourret2015growth} for further
details). In order to reduce the computational cost in the bulk
phases away from the interface, the anisotropy terms and
anti-trapping current are computed only where $|1-\phi^2| \geq
10^{-6}$, i.e. in the vicinity of the solid-liquid interface.
Otherwise, these terms are set to zero.

Homogeneous Neumann (no-flux) boundary conditions are applied on all
domain boundaries for both PDEs. The phase-field
$\psi$ is initialized as the signed distance function to the
liquidus isotherm with positive values in the solid phase. The
dimensionless supersaturation field is initialized based on the
equilibrium concentration over entire domain, i.e. $U = -1$. The
grain index field is initialized to $-1$ in the liquid region
($\psi < 0$) and it is initialized to a Voronoi-based distribution of grain
indices in the solid. To do so, we randomly generate $N$ Voronoi cell centers
in the entire domain using fast Poisson disk sampling \cite{bridson2007fast}
with a random grain index within $[0-89]$.
Grain indices are then allocated to each finite difference grid point
using a classical Voronoi tessellation algorithm.

The simulation domain is sized slightly larger than $l_S\times d_S$, in order to accommodate the entire tail of the melt pool.
In order to calculate a grain map for a solidified length longer than the melt pool length,
we use a standard moving frame algorithm, in which new grid points at the alloy nominal
concentration are added on the right-hand side of the domain, while values of the fields at grid points
leaving the simulation domain on the left-hand side are stored to be later used for reconstructing the grain map for the
entire solidified length (see Figure~\ref{fig:resu:pf1}).

We use $\Delta x = 0.8 W$, and $\Delta x$ is determined
based on a convergence study monitoring the steady-state tip
undercooling as a function of the grid size in unidirectional
solidification considering the
most computationally constraining conditions, namely the (lowest) temperature gradient
and the growth velocity at the tail end of the melt pool.
Under conditions relevant to additive manufacturing, such convergence
study can be quite limiting, since both the dendrite tip radius
and the diffusion length are small (see section~\ref{method:simu:pf}), but it remains critical
if the objective is to quantitatively predict dendrite/cells growth kinetics and resulting grain structures.
Because of this grid size limitation for quantitative predictions,
PF simulations at the scale of the melt pool, even in 2D,
are extremely computationally demanding.
Therefore, advanced acceleration schemes are required, which we briefly mention below.

We implemented the model for parallel computing on
multi-graphic processing units (Multi-GPU) using the computer
unified device architecture (CUDA) programming language.
Even though simulations are massively parallelized, we also aim
at providing a solution that can be implemented on medium-size computing
hardware accessible in-house to most research laboratories or companies.
For this reason, we limit the current study to simulations performed on a single
cluster node with eight Nvidia GPUs (RTX 2080Ti).
A simple layer-wise domain decomposition is used, in which the computational domain
is divided into 8 almost equal layers along the $y$-direction with a
similar (total) number of grid points along $x$-direction. We consider
one extra halo layer of points at the top and bottom rows of each
domain to simplify the imposition of boundary conditions and
inter-GPUs data exchange. The time loop is composed of two
main kernel calls, one for the calculation of the $\psi$-field at
the next time step and one for the calculation of the $U$-field at
the next time step. The time stepping is then achieved by swapping
pointer addresses between arrays containing values of $\psi$ and
$U$ at the current time step and arrays containing values at the
next time step. After the execution of each kernel, the halo grid data
is updated by using direct GPU-GPU communication (memory copy from
one GPU to its neighbor GPUs). In this way, we avoid using
expensive GPU-to-CPU and CPU-to-GPU data transfer.

%=========================================

\subsection{Application to Selective Laser Melting of Inconel 718 alloy}
\label{method:simu}

\subsubsection{Processing conditions}
\label{method:simu:proc}

We illustrate the methods described above and their coupling with the simulation of Selective Laser
Melting of IN718 alloy (Table~\ref{tab:alloy}).  Since we selected a PF model that does not account
for solute trapping, it would not be appropriate to consider a laser scan velocity comparable or
above the onset velocity for solute trapping.  With a typical onset of solute trapping for
solidfication velocities on the order of $\approx 1\,$m/s, we use a scan velocity $V=0.1~$m/s,
which should remain sufficiently below the onset of significant solute trapping.  This assumption is
further discussed in section~\ref{resu:pf:trapping}.  We consider a laser power $P=100~$W, relevant
to actual SLM conditions of Inconel 718, which corresponds to a linear energy density $P/V=1.0$~J/mm
\cite{karayagiz2020finite}.

\subsubsection{Thermal simulations and parameters (FE)}
\label{method:simu:fem}

The parameters used in macroscopic FE simulations are summarized in Table \ref{tab:param:fem}.
Most material properties such as $T_L$, $T_S$, $\rho(T)$, and $c_p(T)$ are calculated using CalPhaD.
The thermal conductivity $\kappa(T)$ is extracted from \cite{mills2002recommended}.
The laser absorption, convection, and emissivity coefficients are chosen according to \cite{zhang2018additive}.
The average powder diameter is selected equal to the layer thickness, and the view factor is chosen according to \cite{sih2004prediction}.

\begin{table*}[!ht]
\centering
\caption{Material properties and process parameters for FE simulations.}
\begin{tabular}{l l l l l}

\hline
& \textbf {Properties} & \textbf{Symbol} & \textbf{Value} &  \textbf{Unit}   \\
\hline
\multirow{6}{10em}{\textbf{Bulk (dense) \\ material}}
&  Solidus temp &  $T_{S}$  &  1554 &  K \\
&  Liquidus temp &  $T_{L}$ &  1625 &  K  \\
&  Boiling temp & $T_V$ &  3038 &  K \\
&  Heat capacity  &  $c_{p}$ &  Fig\ref{fig:calphad}b &  J.kg$^{-1}$.K$^{-1}$ \\
&  Density &  $\rho$ &  Fig\ref{fig:calphad}c &  kg.m$^{-3}$ \\
&  Thermal conductivity &  $\kappa$ &  Fig\ref{fig:calphad}d &  W.m$^{-1}$.K$^{-1}$\\
\hline
\multirow{3}{10em}{\textbf{Powder bed}}
&  Average diameter &  $d$  &  30 &  $\mu$m \\
&  Porosity &  $\xi$ & 0.3 & - \\
&  View factor & $F$ &  0.33 & -\\
\hline
\textbf{Gas (Argon)}
&  Density &  $\rho_g$  &  1.66 &  kg.m$^3$ \\
\hline
\multirow{3}{10em}{\textbf{Substrate \\ (Stainless steel)}}
&  Heat capacity  &  $c_{p}$ &  677 &  J.kg$^{-1}$.K$^{-1}$ \\
&  Density &  $\rho$ &  7900 &  kg.m$^{-3}$ \\
&  Thermal conductivity &  $\kappa$ &  24.9 &  W.m$^{-1}$.K$^{-1}$\\
\hline
\multirow{4}{10em}{\textbf{Process}}
&  Laser Power & $P$ & {100} &  W\\
&  Scan speed &  $V$ & {0.1} &  m.s$^{-1}$\\
&  Beam diameter &  $w$ & { 70} &  $\mu$m\\
&  Thickness of layer & $h_p$ & { 30} &  $\mu$m\\
\hline
\multirow{4}{8em}{\textbf {Boundary conditions}}
&  Absorption coefficient & $A$ &  0.55 & - \\
&  Convection coefficient &  $h_c$  &  15 &  W.m$^{-2}$.K$^{-1}$ \\
&  Emissivity &  $\epsilon$ &  0.3 & -\\
&  External temperature &  $T_{\rm ext}$ &  273 &  K\\
\hline
\end{tabular}
\label{tab:param:fem}
\end{table*}

\begin{table*}[t!]%
\centering
\caption{Material properties and process parameters for PF simulations.}
\label{tab:param:pf}%
\begin{tabular}{l l l l}%
\hline%
\textbf{Properties} & \textbf{Symbol} & \textbf{Value} & \textbf{Unit} \\ %
\hline%
 Nominal alloy concentration & $c_\infty$ & 5.0 & wt\%\,Nb \\%
 Solute partition coefficient & $k$ & 0.37 & - \\%
 Liquidus slope & $m$ & 9.0 & K.wt\%\,Nb$^{-1}$\\%
 Liquid diffusion coefficient & $D_l$ & $2.44\times 10^{-9}$ & m$^2$.s$^{-1}$\\%
 Pseudo-binary solvent melting temperature & $T_M'$ & 1670  & K\\%
 Pseudo-binary reference temperature (Solidus) & $T_0$ & 1549  & K\\%
 Gibbs-Thomson coefficient & $\Gamma$ & $2.49\times 10^{-7}$  & K.m\\%
 Interface anisotropy & $\varepsilon_4$ & 0.02  & - \\%
 Grid element size & $\Delta x$ & 5& nm \\
\hline%
\end{tabular}
\end{table*}

We ensured the numerical convergence of the simulation, with respect to the grid element size, the
number of deposited layers, and the domain size, by monitoring the dimensions of the melt pool
($l_S$, $l_L$, $d_S$, and $d_L$).  We found that, in order to obtain a steady melt pool size and
shape, we needed to consider (1) a minimum of two grid elements within the powder layer thickness
$h_p=30~\mu$m, and (2) at least five or six powder layers on top of the substrate considering
successive heating and cooling stages, using a 0.166~s cooling time between layers, which was found
to be sufficient for the part to cool down close to the external temperature.  We also identified
that a domain size of $0.21\times1.2\times0.6$~mm$^3$ was sufficient to stabilize a steady melt
pool.  Hence, we simulated ten powder layers using a grid element size of 15~$\mu$m and used the
steady temperature profile within the tenth layer as thermal field for the PF simulations.

Moreover, in order to study the effect of the temperature-dependent properties, we performed
additional simulations with either constant conductivity or constant density, all other parameters
remaining the same.  For these simulations, the conductivity or density of the dense material was
assessed at the liquidus temperature as 29\,W.m$^{-1}$.K$^{-1}$ or 7400\,kg.m$^{-3}$, respectively.
We also performed one additional simulation in which the thermal properties of the powder bed were set
equal to that of the bulk (dense) material in order to estimate the effect of the difference in
properties between powder bed and dense material.

\subsubsection{Microstructure simulations and parameters (PF)}
\label{method:simu:pf}

\begin{figure*} [!t]
    \centering
    \includegraphics[width=1.8\columnwidth]{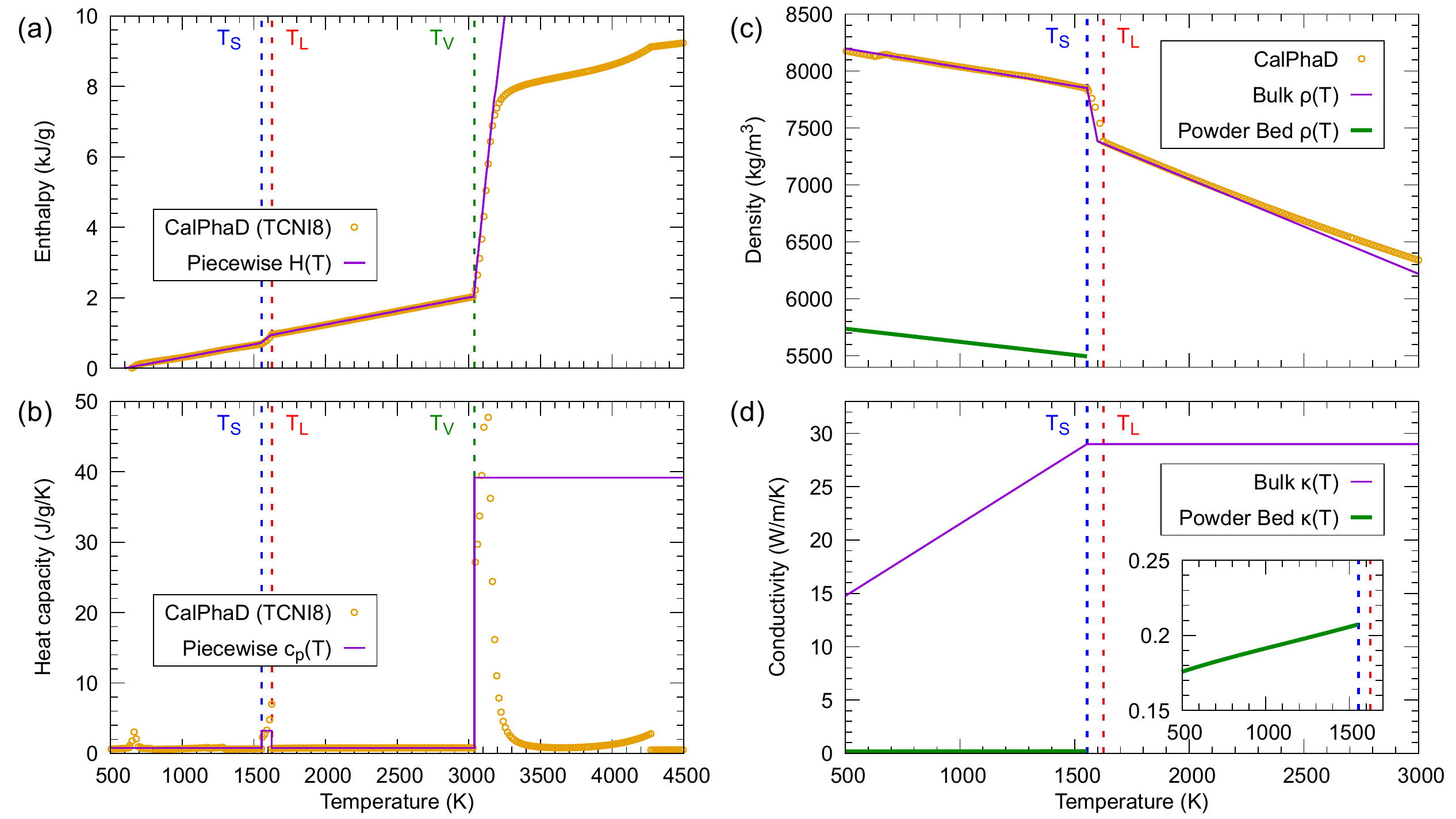}
    \caption{
    Temperature-dependent material properties for alloy IN718: (a) Enthalpy, (b) Heat capacity, (c) Density, (d) Conductivity \cite{mills2002recommended}.
    CalPhaD calculated data (database TCNI8) appear as symbols while piecewise linear fits used in FE simulation appear as lines (thin purple for bulk, thick green for powder bed.)
	}
    \label{fig:calphad}
\end{figure*}

Table \ref{tab:param:pf} shows the physical and computational parameters used in PF simulations.  As
explained in Section~\ref{method:calphad}, the solute (Nb) partition coefficient and liquidus slope
were calculated considering the full IN718 alloy (Table~\ref{tab:alloy}), and the corresponding
pseudo-binary $T_M'$ and $T_0\equiv T_S'$ used in the PF model were calculated to match the liquidus temperature
$T_L$ of the full alloy with that of the pseudo-binary approximation.  Notably, the solidus
temperature considered in the pseudo-binary PF simulation ($T_S'=1549~$K) remains close to the
solidus of the full alloy calculated using CalPhaD and considered in the FE simulation
($T_S=1554~$K).  Using ab initio molecular dynamics simulations, Walbr\"uhl and collaborators have
estimated the diffusion coefficient of Nb in Ni (Ni-10at.\%Nb) between 1903 and 2303 K, and assessed
Arrhenius prefactor $D_0\approx1.22\times 10^{-7}~$m$^2$/s and activation energy
$E\approx55.3$~kJ.mol$^{-1}$.K$^{-1}$ \cite{walbruhl2018atomic}.  We use this expression to estimate
the diffusion coefficient in the vicinity of 1700~K, which we use as constant in the PF simulation
with $D\approx2.44\times10^{-9}~$m$^2$/s.  The Gibbs-Thomson coefficient of the solid-liquid
interface is calculated as $\Gamma=\gamma_0T_M/L\approx2.49\times10^{-7}~$K.m, using 
the melting temperature of pure Ni, $T_M=1728~$K, and the latent heat of fusion, $L=2.08\times10^9~$J.m$^{-3}$,
calculated from the ThermoCalc (TCNI8) calculation of $h(T)$ for pure Ni, and an
interface excess free energy $\gamma_0\approx0.3~$J.m$^{-2}$ consistent with those calculated for
pure Ni with molecular dynamics (capillary fluctuation method) in several references (from 0.27 to
0.36~J.m$^{-2}$ in Refs~\cite{hoyt2003atomistic, jiang2008size, asadi2015two}).  We use an
anisotropy strength for the interface excess free energy of $\varepsilon_4=0.02$, which is of the
same order as identified by these atomistic simulations (e.g. $\varepsilon_4\approx0.018$ in
\cite{hoyt2003atomistic}), considering that here we only use the fourfold anisotropy component according to
Eq.~\eqref{eq:ast}.

For these parameters, a thorough convergence analysis (see sections~\ref{method:pf:implem} and
\ref{resu:pf:conv}) revealed that a grid spacing $\Delta x=5$~nm was necessary for well-converged
simulations.  As discussed later in section~\ref{resu:pf:scales}, this value is consistent with the
relevant length scales under these conditions.  Hence, considering a simulation domain size of
$250\,\mu$m$\,\times\, 100\,\mu$m, a simulated time of 5\,ms, and a stable time step $\Delta
t=0.76$\,ns, this resulted in 50\,000\,$\times$\,20\,000 grid points (i.e. over 2 billion degrees of
freedom) and about 6.6 million iterations.  While this is arguably a large simulation, it was
nonetheless achievable in a reasonable time (under 10 days) with a reasonable computing hardware
(one compute node equipped with eight Nvidia RTX 2080Ti GPUs).

%=================================================================

\section{Results and Discussion}
\label{resu}

\subsection{Temperature-dependent thermophysical properties}
\label{resu:calphad}

In the first step of our methodology, the CalPhaD method is used to calculate the
temperature-dependent properties of IN718, namely enthalpy, heat capacity, and density.
Figure~\ref{fig:calphad} shows the resulting CalPhaD results (orange symbols), as well as the
resulting piecewise linear approximations for $h(T)$ (i.e. piecewise constant $c_p(T)$) and
$\rho(T)$ used in the FE simulations (purple solid lines).  The temperature-dependent conductivity
(Figure~\ref{fig:calphad}d) was extracted from \cite{mills2002recommended}.  The powder bed density and
conductivity (thick green lines) are estimated using
Eqs~\eqref{eq:powderrho}-\eqref{eq:powderkappa}.

%=========================================

\subsection{Macroscopic thermal field}
\label{resu:fem}

\begin{figure} [b!]
    \centering
    \includegraphics[width=.95\columnwidth]{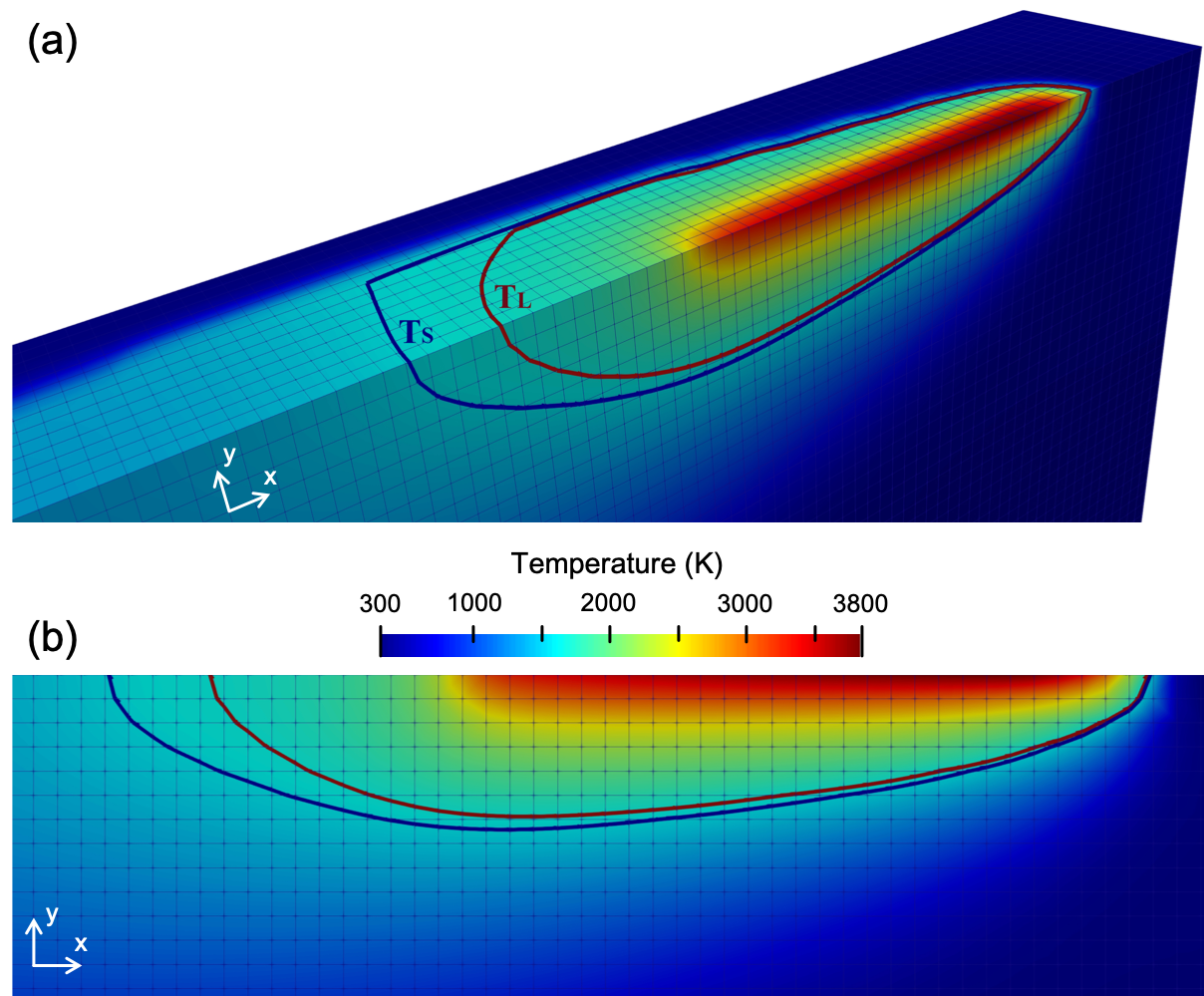}
    \caption{Temperature distribution during heating stage of tenth layer as predicted by finite element simulation.}
    \label{fig:resu:fem3d}
\end{figure}

Figure \ref{fig:resu:fem3d} shows the temperature field during the heating stage of the tenth layer from the FE thermal simulation. The solidus and liquidus isotherms appear as blue and red lines, respectively.
The resulting melt pool dimensions are  $l_L \approx 185~\mu$m, $l_S\approx 249~\mu$m, $d_L\approx 88~\mu$m, and $d_S\approx96~\mu$m.
Note that in the PF simulations, we used approximate dimensions  $l_L=190 ~\mu$m, $l_S=245~\mu$m, $d_L=90~\mu$m, and $d_S=95~\mu$m.

\begin{figure} [b!]
    \centering
    \includegraphics[width=.95\columnwidth]{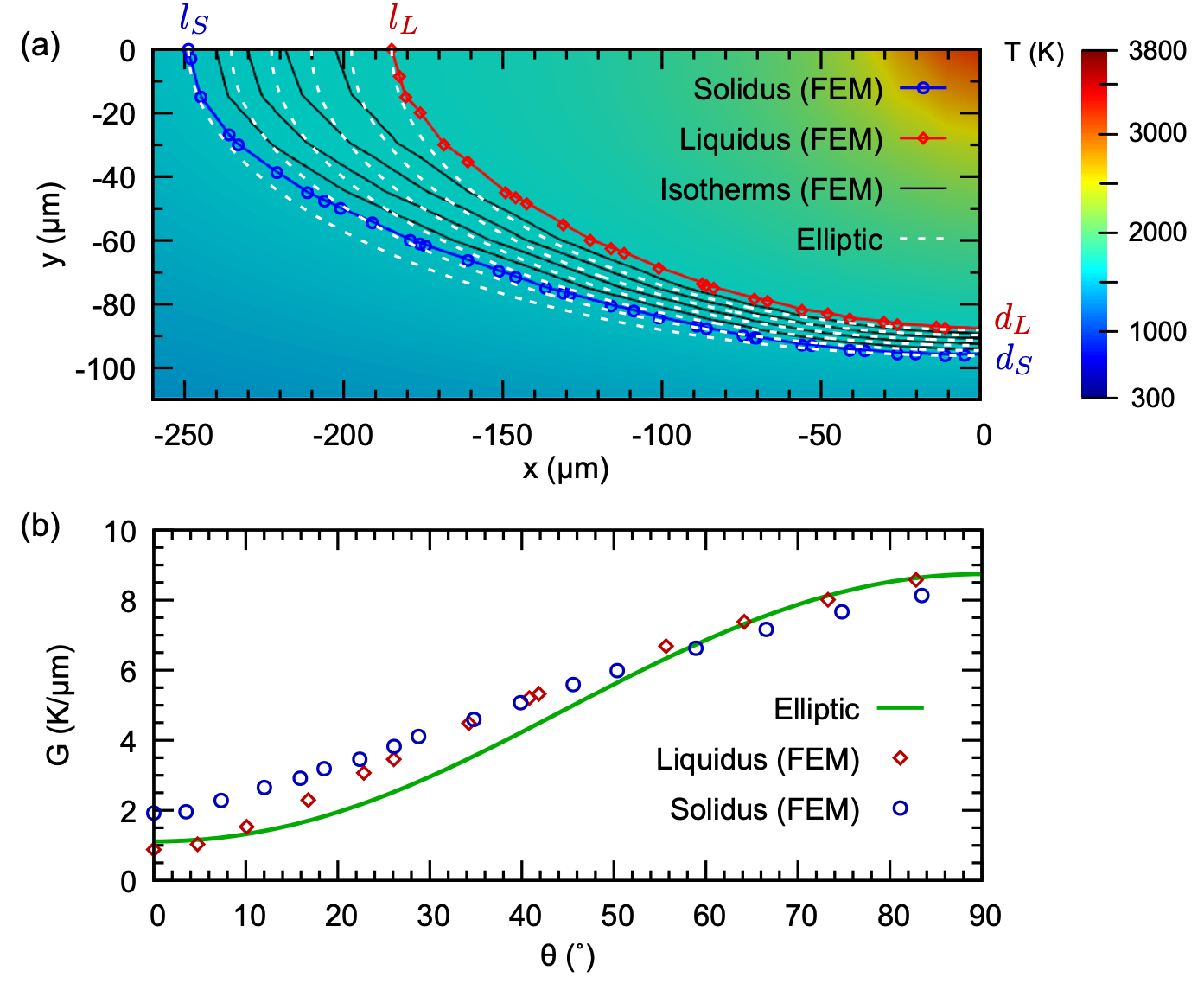}
    \caption{Melt pool shape within the longitudinal section of Figure~\ref{fig:resu:fem3d}b,
comparing FE results and the elliptical approximation (Eq.~\eqref{eq:tellip}): (a) FE-predicted
temperature field (color background) and isotherms for $T=T_S=1554\,$K (solid blue line and
symbols), $T=T_L=1625\,$K (solid red line and symbols) and $T=1569$, 1583, 1597, and 1611\,K (black
solid lines), compared to elliptic approximation along the same temperatures (white dashed lines);
(b) temperature gradient as a function of the polar angle $\theta$ (see Figure~\ref{fig:elliptic}) as
predicted by FE along the solidus (blue circle symbols) and liquidus (red diamond symbols) isotherms
compared to the elliptic approximation (solid green line).}
    \label{fig:resu:meltpool}
\end{figure}

Figure~\ref{fig:resu:meltpool}a compares the tail half of the melt pool within the central
longitudinal section (solid lines and symbols) with the elliptic approximation using
Eqs~\eqref{eq:rl}-\eqref{eq:rs} and used as input in the PF simulations (white dashed lines).
Figure~\ref{fig:resu:meltpool}b compares the resulting temperature gradient as a function of the
angle from the top of the domain, $\theta$, as calculated in the FE simulations along $T_S$ and $T_L$
isotherms (symbols) and using the elliptic approximation (solid green line).  Using the elliptic
approximation, the polar component of the temperature gradient, $(1/r)\partial T/\partial \theta$,
is the only term distinguishing the gradient measured along the solidus or along the liquidus line.
Yet, its magnitude remains below $10^4\,$K/m, which is negligible compared to the radial component,
$\partial T/\partial r$, which is of order $10^6$ to $10^7\,$K/m.  The curve in
Figure~\ref{fig:resu:meltpool}b uses all terms along the $T_L$ isotherm, but the plot along $T_S$ or
the plot considering only the $\partial T/\partial r$ term are virtually undistinguishable.
Consequently, for an elliptically shaped melt pool, a reasonable estimation of the temperature
gradient can be conveniently obtained directly for the polar angle as $\partial T/\partial r =
(T_L-T_S)/(r_L(\theta)-r_S(\theta))$, using $r_L(\theta)$ and $r_S(\theta)$ from
Eqs~\eqref{eq:rl}-\eqref{eq:rs}.  As shown in Figure~\ref{fig:resu:meltpool}, in spite of a small
deviation, the analytical function provides a reasonable approximation of the temperature field
between $T_L$ and $T_S$.

Since the growth velocity $V_{\rm gr}$ is at most equal to $V=0.1~$m/s and the temperature gradient
is between $10^6$ and $10^7~$K/m, the ratio $G^2/V_{\rm gr}$ is higher than
$10^{13}\,$K$^2$s/m$^{3}$, which should be sufficiently high to lead to fully epitaxial growth of
columnar microstructure, consistently with the assumption made in the PF simulation.  As a
comparison, Knapp et al. \cite{knapp2019experiments} estimated the condition for a fully columnar structures for any $G^2/V_{\rm
gr}$ above $1.52\times10^{11}\,$K$^2$s/m$^3$ for Inconel\,718 alloy, however considering electron
beam melting.

\begin{figure} [b!]
    \centering
    \includegraphics[width=.95\columnwidth]{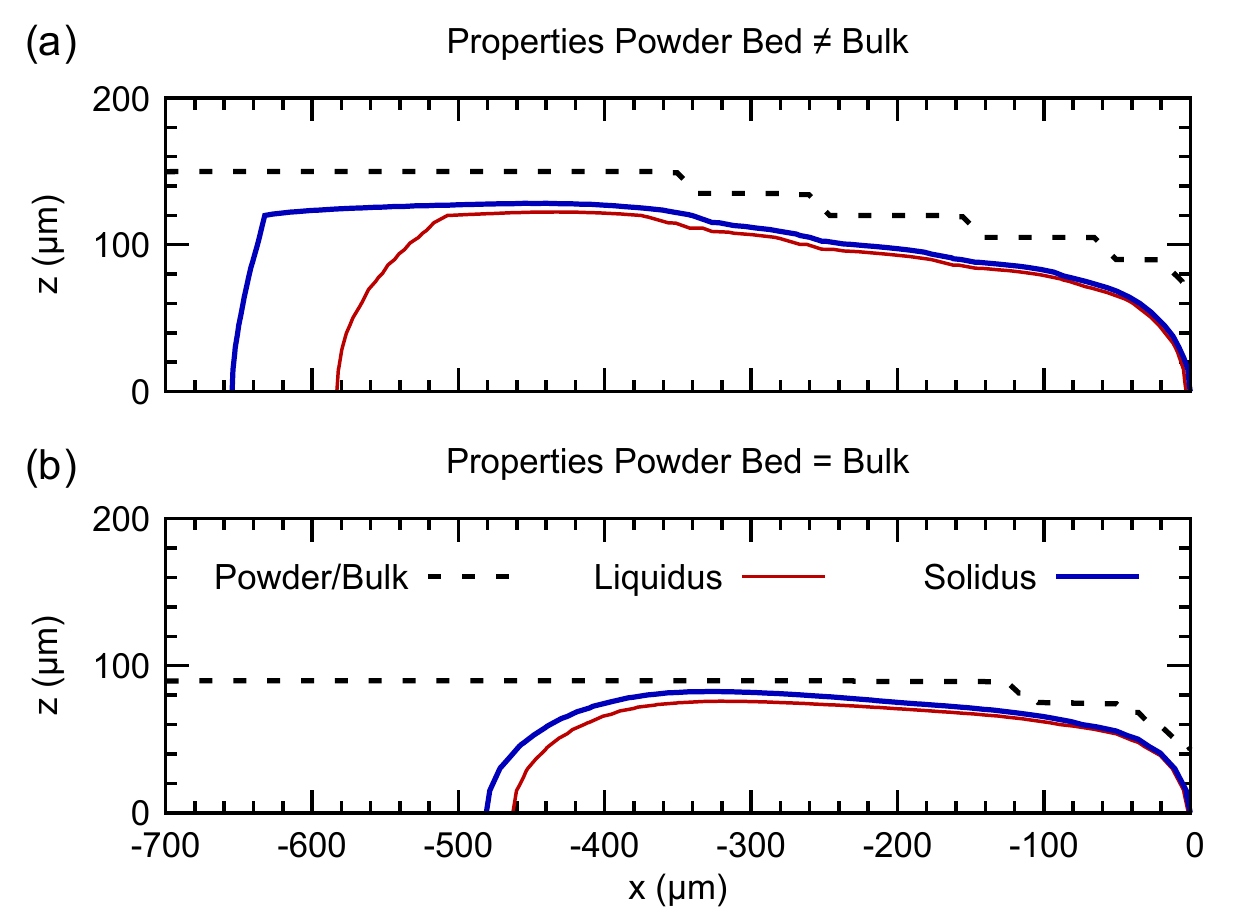}
    \caption{Top view of the liquidus and solidus isotherms (solid lines) and computed boundary
between powder bed and bulk (dense) states (dashed line) when considering different thermal properties
in dense and powder bed states (a), and when the powder bed has the same properties as the bulk
material (b).}
    \label{fig:resu:meltpooltop}
\end{figure}

The thermal field in Figure~\ref{fig:resu:fem3d}a exhibits, on the top surface, a kink in both
isotherms, most prominently for the solidus, in the tail end of the melt pool.  This feature is due
to the presence of a straight boundary between the bulk (dense) regions directly below and behind
the laser path, while the material on the side of the path is still in the powder state, hence with
a significantly lower density and even more importantly a much lower conductivity (see
Figure~\ref{fig:calphad}).  Figure~\ref{fig:resu:meltpooltop} illustrates the top surface isotherms
(solid lines), as well as the computed boundary  between powder bed and dense states (dashed line) for
the current simulation (a) as well as in a simulation in which the entire domain has the
thermophysical properties of the dense material (b).  Not only does the kink in both isotherm
disappear, but the melt pool size is also significantly reduced in the latter case, due to the
easier flow of heat along the sides made of dense more conductive material.

\begin{figure} [b!]
    \centering
    \includegraphics[width=.95\columnwidth]{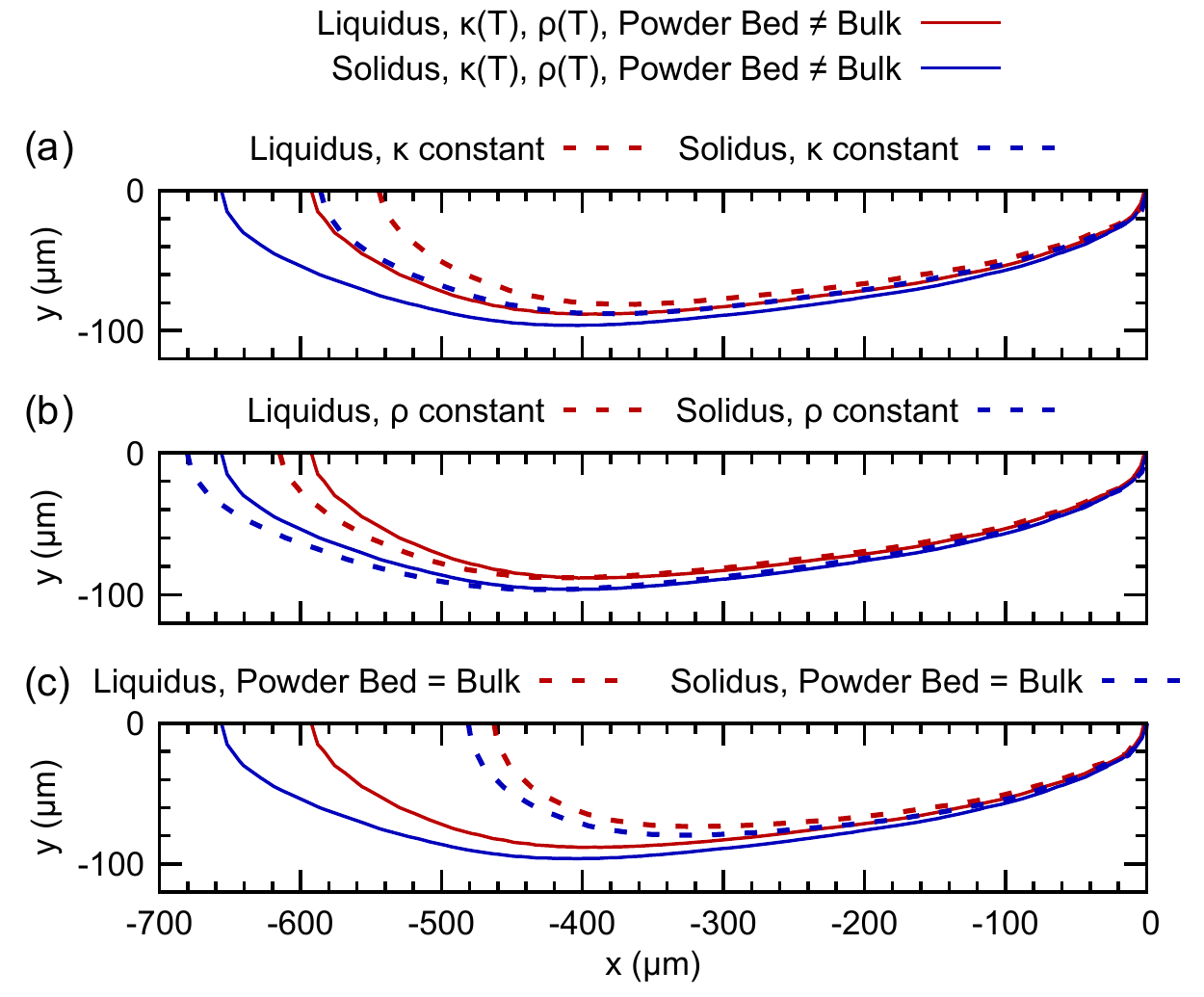}
    \caption{Liquidus (red) and solidus (blue) isotherms in the central longitudinal section for the
reference simulation of Figure~\ref{fig:resu:fem3d} (solid lines) compared to equivalent simulations
(dashed lines) with a constant conductivity (a), constant density (b), or equal properties in the powder bed
and bulk material (c).}
    \label{fig:resu:constantparams}
\end{figure}

Finally, we also assess the effect of temperature-dependent properties on the resulting melt pool
shape (liquidus and solidus isotherms) in the longitudinal section.
Figure~\ref{fig:resu:constantparams} shows the reference results of Figure~\ref{fig:resu:fem3d} (solid
lines) compared to equivalent simulations (dashed lines) considering a constant conductivity (a),
constant density (b), as well as equal properties in the powder bed as in the bulk material (c). A
constant conductivity tends to reduce the size of the melt pool, while a constant density tends to
slightly increase it.  However, the most important effect appears to be the consideration of
different powder bed and bulk properties, as seen in Figure~\ref{fig:resu:constantparams}c.  This was
already seen in Figure~\ref{fig:resu:meltpooltop}, and further highlights the importance of accounting
for the thermal properties of the powder bed in order to obtain reliable thermal simulations of
powder-bed fusion processes.

%=========================================

\subsection{Microstructure growth in the melt pool}
\label{resu:pf}

\subsubsection{Convergence analysis}
\label{resu:pf:conv}

The numerical convergence analysis of our PF simulations was performed on a reduced relevant problem involving a one-dimensional thermal field, i.e. using the classical frozen temperature approximation, with a pulling velocity equal to $V=0.1~$m/s and a temperature gradient $G=10^7~$K/m.
Measuring the steady state dendrite tip undercooling achieved for different grid element sizes we found that results started deviating substantially for $\Delta x\approx5\,$nm or higher, thus identifying the grid spacing necessary to achieve quantitative simulations.
Increasing the grid spacing any further promotes the interaction between neighbor dendrites and leads to the formation of pockets of highly segregated liquid between them, which resemble patterns observed in rapid solidification experiments \cite{keller2017application, boettinger1988formation} but tend to disappear when the discretization is refined.

\subsubsection{Relevant length scales}
\label{resu:pf:scales}

The value of $\Delta x\approx5\,$nm is consistent with the most important physical length scales in
the melt pool solidification problem.  Indeed, for $\epsilon_4=0.02$, the two-dimensional one-sided
tip selection parameter is expected to be $\sigma^*=2Dd_0^*/(R^2V)\approx0.15$ (see Figure\,1 and
Eq.\,(4.3) in \cite{barbieri1989predictions}).  With the alloy parameters in
Table~\ref{tab:param:pf}, an approximate estimation of the steady-state growth leads to a steady
dendrite tip radius $R\approx39.3\,$nm, with a capillary length $d_0^*\approx4.75\,$nm at a
dimensionless tip undercooling $\Delta\approx0.497$, i.e. a P\'eclet number $P=RV/(2D)\approx0.806$
(whereas $d_0$ at the solidus temperature $T_0$ is close to 3.25\,nm), and a diffusion length
$D/V=24.4\,$nm.  Therefore, the grid element size $\Delta x=5\,$nm is only eight times smaller than
the steady tip radius, five times smaller than the steady diffusion length, and of the same order as
the capillarity length at the tip.  While only an approximate order-of-magnitude analysis (none of
the actual growth in the melt ever really reaching steady state, and the laser velocity $V$ being
only relevant to the tail region of the melt pool), this still provides a sensible picture of why
the grid element size cannot be taken any coarser without compromising accuracy.

\begin{figure*} [t!]
    \centering
    \includegraphics[width=.95\textwidth]{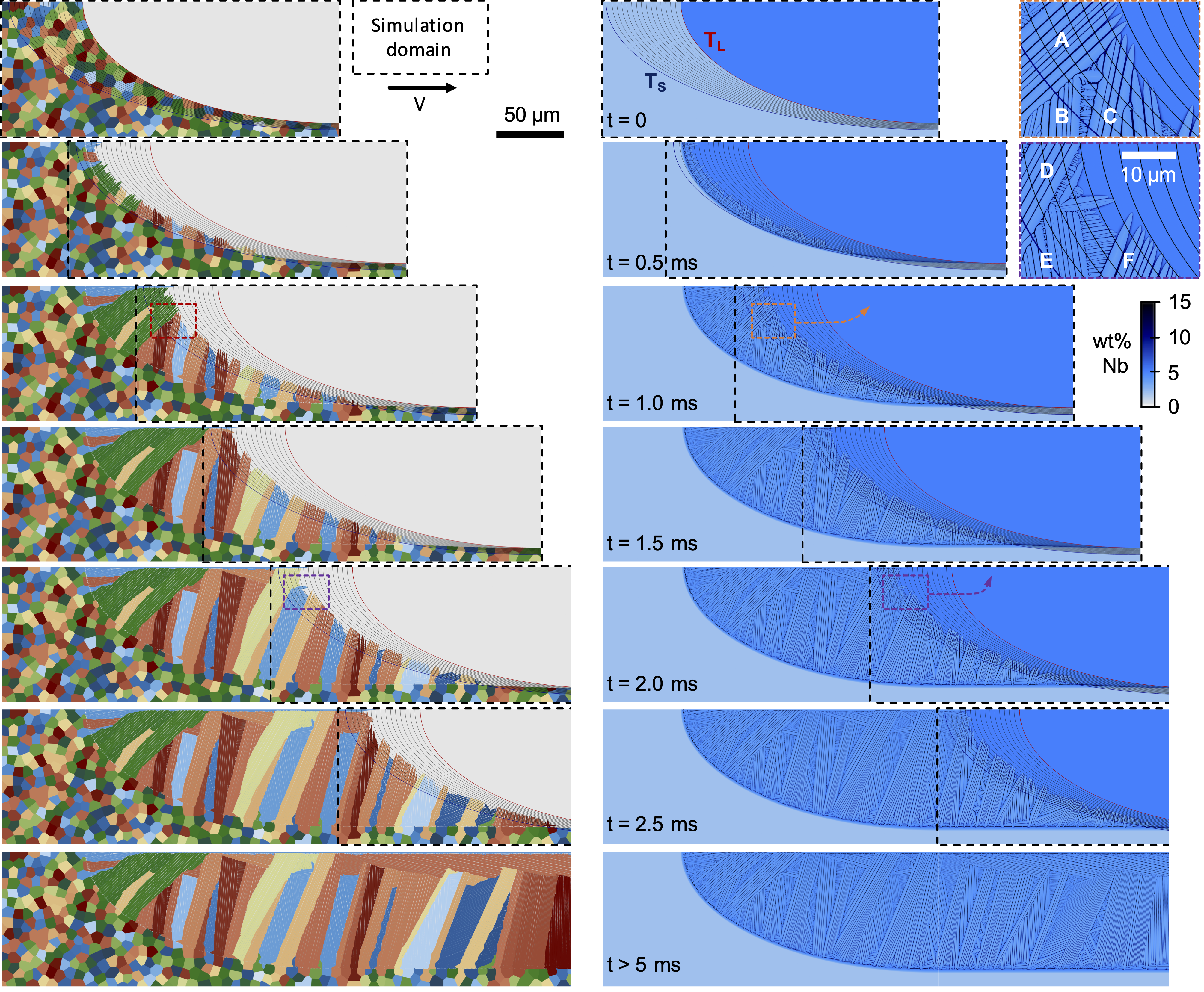}
    \caption{ Phase-field simulation results showing grain structure formation (left) and solute
(Nb) concentration field (right). Iso-temperature lines show $T=T_L=1625~$K (red),
$T=1550~$K$~\approx T_S$ (blue), and intermediate temperature with steps of 5~K (black). The simulation
domain, moving at a velocity $V$, is delimited with dashed black lines.  Zoomed-in regions at the
top right are marked with dashed rectangle in the resulting full-scale maps at $t=1.0\,$ms and
$t=2.0\,$ms.
}
    \label{fig:resu:pf1}
\end{figure*}

\subsubsection{Solute trapping}
\label{resu:pf:trapping}

Finally, we discuss the assumption of solid-liquid interface equilibrium with respect to solute trapping, and whether a laser velocity $V=0.1\,$m/s is sufficiently low for this assumption to remain valid.
According to the continuous growth model (CGM) \cite{aziz1982model, aziz1988continuous}, solute partitioning at the interface changes with the interface velocity $V_i$ like $k(V_i)=[k_e+V_i/V_D]/[1+V_i/V_D]$, with $k_e$ the equilibrium partition coefficient and $V_D$ the solute diffusion velocity through the interface.
Using the parameters of the pseudo-binary alloy (Table~\ref{tab:param:pf}) and an order of magnitude for the physical interface width $l_A\approx 1\,$nm, one can approximate the diffusion velocity as $V_D\approx1.44~$m/s (see Eq.\,(63) in \cite{ahmad1998solute}).
Therefore, the considered laser velocity $V=0.1\,$m/s, which is the highest growth velocity experienced in the melt pool at its tail end, seems to be sufficiently lower than $V_D$ for solute trapping to remain negligible.
However, even though we consider $V\ll V_D$, the resulting change of partition coefficient from the CGM is just above 10\%, which remains small but could become important at larger $V$.

The diffuse interface width used here is $W=6.25\,$nm, which is sensibly higher than the actual width of the solid-liquid interface. Should the required value of $W$ be further reduced for convergence, e.g. for higher $V$, it is worth noting that using physically realistic diffuse interface width can lead to prediction of solute trapping effect in good agreement with the CGM \cite{ahmad1998solute}.

\subsubsection{Full melt pool simulations}
\label{resu:pf:meltpool}

The results of the 2D PF simulation of solidification at the full melt pool scale appear in Figure~\ref{fig:resu:pf1}, showing the time-evolution (top to bottom) of the grain structure (left) and solute (Nb) concentration field (right).
The grain map illustrates the growth competition at the melt pool scale, while the solute map gives a more detailed insight into the dendritic structures within the grains.
While this sole two-dimensional simulation is not sufficient to draw statistically-relevant conclusions on grain growth competition in AM-relevant conditions, trends appear that highlight interesting similarities and differences with traditional, Bridgman-like, directional solidification (DS).

\begin{figure*} [t!]
    \centering
    \includegraphics[width=.95\textwidth]{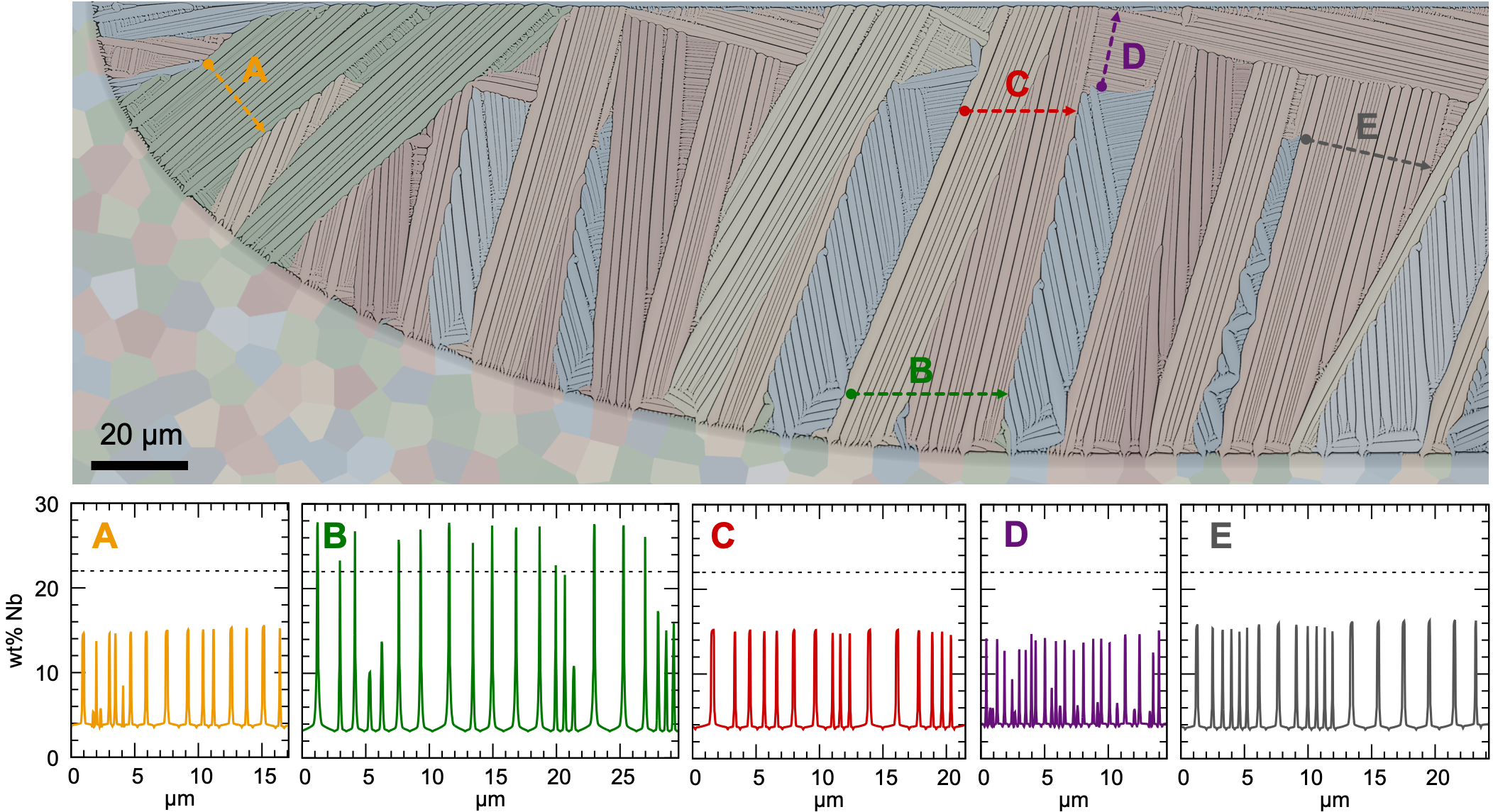}
    \caption{Solute (Nb) segregation profiles along different line scans (A--E) in the solidified region as predicted by phase-field simulations.
    The dashed line in the bottom plots marks the solute concentration of the eutectic point (L$\rightarrow$Ni$_{\rm fcc}$+Ni$_3$Nb) in the Ni-Nb phase diagram.}
    \label{fig:resu:pf2}
\end{figure*}

Similarly to DS, dendritic elimination (impingement) and sidebranching events are responsible for the orientation selection of converging and diverging grain boundaries, respectively \cite{tourret2015growth, tourret2017grain} (see, e.g., top right zoomed-in regions in Figure~\ref{fig:resu:pf1}, each highlighting the grain growth competition of three grains).
However, in contrast with DS, the amplitude and direction of the temperature gradient, as well as the local cooling rate are constantly changing in the vicinity of each nascent grain boundary (GB).
This makes it nontrivial to identify with absolute certainty the ``favorably'' or ``unfavorably'' oriented grains forming the GB, as these roles may switch during the process.

Two interesting observations can be readily made from these results.
First, most grains selected by the growth competition are slightly tilted forward with respect to the vertical direction.
For most grains, this corresponds to a principal dendritic growth direction, i.e. a main crystalline orientation.
However, some larger grains manage to prevail through successive sidebranching in spite of a substantial misorientation of their crystalline dendritic axes with this ``mesoscopic'' direction of maximum elongation of the grain (see bottom right panel in Figure~\ref{fig:resu:pf1}).
The prevalence of tilted columnar grains, regardless of their inner crystalline orientation, highlights the importance of the simulation at the full melt pool scale, as these would not naturally emerge from grain growth competition in a reduced subset of the melt pool.

Second, a noticeable range of different primary dendritic spacing (PDAS) range emerges (see, e.g., final, bottom-right, dendritic structure in Figure~\ref{fig:resu:pf1}).
This PDAS heterogeneity occurs not only among different grains but also within a same grain.
Such examples appear in the zoomed-in areas (top right) of Figure~\ref{fig:resu:pf1}, most notably within grain A and grain E.
Grain A exhibits a heterogeneity among spacings that emerged directly from the initial almost-planar (slightly curved) interface destabilization.
Such heterogeneity may be attributed to the quasi-steady, yet not quite steady, growth conditions, combined with the fact that spacing homogenization within a grain can take substantially longer in time than it takes to reach a steady or quasi-steady growth velocity and undercooling \cite{clarke2017microstructure}.
On the other hand, the locally smaller microstructural length scale in grain E is due to the fact that the lower spacing region emerges from sidebranching, thus forming a region with locally secondary dendrite arm spacings (SDAS) along the diverging GB.

Finally, we illustrate in Figure~\ref{fig:resu:pf2} a potential use of these results in terms of digital microstructure characterization.
There, we show the grain structure (color background) overlayed with the Nb concentration map (gray level), as well as line scans of the Nb concentration in different regions of the melt pool (bottom plots).
Such signals deserve two important remarks.
First, periodicity of the signals gives the average primary spacing within the grain, which could be conveniently extracted in a systematic manner using adapted spectral filtering techniques.
Second, the extent of interdendritic Nb segregation allows identifying the region in which the secondary phases are most prone to form.
Since this composition field is reconstructed from still partially liquid regions (due to the moving frame algorithm), it is appropriate to compare these segregation peaks to the eutectic triple point (L$\rightarrow$Ni$_{\rm fcc}$+Ni$_3$Nb) at $T\approx1295\,^\circ$C and $c\approx21\,$wt\%Nb in the Ni-Nb phase diagram (according to ThermoCalc TCNI8 calculations), marked with a dashed line in the bottom plots of Figure~\ref{fig:resu:pf2}.
Therefore, this analysis suggests that regions at the bottom of the melt pool, such as region B, are most prone to the formation of intermetallic Ni$_3$Nb phase.

%=================================================================

\section{Summary and Perspectives}
\label{summary}

In this article, we presented a multiscale modeling framework for the simulation of powder-bed fusion of metallic alloys.
The framework combines and couples the following methods:
\begin{itemize}
\item CalPhaD calculation of temperature-dependent properties and phase diagram, thus allowing the investigation of alloy chemistry;
\item Three-dimensional finite element thermal simulation of laser melting, considering distinct properties in distinct regions and CalPhaD-based temperature-dependent properties;
\item Two-dimensional phase-field simulations of microstructure development by polycrystalline solidification in the melt pool.
\end{itemize}

We applied the methodology to simulate selective laser melting of Inconel 718 superalloy.
We discussed the effect of temperature-dependent parameters and the importance applying distinct properties in the powder bed and dense regions for the prediction of the melt pool size and shape.
Finally, we simulated the dynamical selection of grain structure through polycrystalline growth competition using 2D quantitative simulations at the scale of the entire melt pool, highlighting some key similarities but also differences with equivalent simulations typically performed on a reduced subset of the melt pool.

This study arguably constitutes an important step forward in the context of Integrated Computational materials Engineering (ICME) for powder-bed fusion processes.
However, it also contains a number of limitations, most of which relate to ongoing work and future directions. 

Regarding macroscopic simulations, the next step is a coupling with thermomechanics, including fluid dynamics, and plasticity.
Simulation of fluid flow would allow predicting defect formation \cite{khairallah2016laser, panwisawas2017keyhole, yan2017multi, zhang2018numerical, tang2020physics}, but it would also permit extending the approach to powder-bed melting in keyhole mode. 
If a similar level of accuracy is sought in the modeling of dendritic growth, a complete two-way coupling between solidification and fluid flow in the liquid would likely require a multiscale approach (e.g. concurrent grids or methods), a computationally efficient and scalable technique for the modeling of the flow (e.g. Lattice Boltzmann Method \cite{sakane2017multi, takaki2020large,yamanaka2021multi}), and/or advanced algorithms for code acceleration via parallelization \cite{shimokawabe2011peta, sakane2017multi} and/or adaptive meshing \cite{greenwood2018quantitative,zhang2018macroscopic, sakane2022parallel}.
Alternatively, one may also conceive a one-way coupling strategy by imposing temperature and solute fields calculated via macroscopic simulations at a distance --- larger than the diffusive boundary layer yet smaller than the typical hydrodynamic length --- ahead of the solidification front in microscale simulations. 
Ongoing extension to thermomechanics simulations in the solid state will also allow the prediction of important features related to the print quality, such as residual stresses and part distortion \cite{zhang2018numerical}.
The level of details required to accurately predict the melt pool shape may prevent the simulation of entire components, unless leveraging advanced numerical strategies (e.g. adaptive meshing \cite{zhang2018macroscopic, baiges2021adaptive}).
However, we trust that the current physics-based approach should be scalable for the simulation of ``mesoscopic'' representative volume elements.

In terms of the microstructure PF simulations, main limitations relate to the pseudo-binary alloy approximation, the lack of solute trapping, and the absence of solid-state microstructure evolution. The extension to multicomponent alloys or solute trapping will require the use of dedicated models (e.g. \cite{eiken2006multiphase, nestler2011phase, ohno2012quantitative, zhang2012phase, kundin2015phase}).
Notably, approximate yet pragmatic extensions of the current model were recently proposed that allow some amount of solute trapping matching CGM theory at growth velocity close to $V_D$ \cite{pinomaa2019quantitative, kavousi2021quantitative}.
Solid-state microstructure evolution during heat treatments, either intrinsic (e.g. in the heat affected zone) or extrinsic (e.g. via ageing), could be included asynchronously using dedicated phase-field models (e.g. \cite{wen2003phase, zhu2004three,  boussinot2009phase, cottura2012phase, ali2020role}).
Another important aspect to include is nucleation, since its rate determines the extent of columnar/equiaxed grain structures.
Its introduction is rather straightforward using phenomenological approaches (e.g. randomly seeding nuclei) \cite{granasy2019phase}.
However, this would also introduce additional parameters (e.g. nuclei density and activation undercooling) which would need to be carefully calibrated to yield reliable predictions.

Ultimately, the resulting microstructure analysis proposed here remains semi-quantitative, mostly due to the two dimensional simulations, the pseudo-binary alloy approximation, and the fact that a statistical (high-throughput) exploration would be required to extract statistically meaningful trends and conclusions.
However, we trust that this type of methodology offers a promising path forward for ICME in the context of alloy design and process optimization in fusion-based AM of metals.
The level of microstructural details, the limited number of calibration parameters, and the fact that such simulation is achievable using reasonable computing resources shall open the way to high-throughput statistical analyses, which will contribute to tackle the pervasive issues of uncertainty and reproducibility in metal AM.

\section*{Acknowledgements}

This investigation was supported by the Spanish Ministry of Science under the {\it Retos-Colaboraci\'on} project ENVIDIA (Ref. RTC-2017-6150-4).
D.T. also gratefully acknowledges support from the Spanish Ministry of Science through a Ram\'on y Cajal Fellowship (Ref. RYC2019-028233-I).

\bibliography{References}

\end{document}